# Soft Error Rate in Space:
# A Unified Analytical Approach

G. I. Zebrev, N. N. Samotaev, R. G. Useinov, A. A. Mateiko, A. S. Rodin

[1] Micro- and Nanoelectronics Department, National Research Nuclear University MEPHI (Moscow Engineering Physics Institute), Moscow, Russia; gizebrev@mephi.ru (G.Z.); nnsamotaev@mephi.ru (N.S.);

**Abstract:** A new physics-based model for analytical calculation of Soft Error Rate (SER) in digital memory circuits under the influence of heavy ions in space orbits is proposed. This method is based on parameters that are uniquely determined from the results of ground tests under normal ion incidence. It is shown that preliminary averaging over the total solid angle within the standard inverse cosine model allows one to take into account the effect of isotropic flow, which increases the effective SER. The model includes the ability to estimate the contribution to SER of the low LET spectrum region, which is very important for modern ICs with low Single Event Upset tolerance.

## 1. Introduction

As reliance on space-based digital electronics grows, understanding and mitigating radiation-induced failures becomes paramount. One of the major challenges facing these systems in the harsh environment of space is transient errors arise from the ionization of separate energetic particles, particularly heavy ions, leading to bit-flips in memories and logic of digital systems known as Single Event Upsets (SEUs) or soft errors. [1] Accurate calculation of soft error rates (SERs) is essential for the design and reliability assessment of such systems [2].

Traditionally, SER prediction has involved complex simulation and modeling techniques, which often rely on empirical data and specific material responses to radiation. In practice, analytical models are often used for this purpose, which have advantages over cumbersome numerical calculations in terms of convenience and have approximately the same accuracy class. For example, a recent review lists and compares at least 12 analytical methods for estimating SER [3].

This paper presents an analytical methodology to predict SER in space-based digital electronics, integrating various important aspects, including angular dependence of SEU cross sections and contribution of low LET portion of Linear Energy Transfer (LET) spectra.

## 2. Methodology of Soft Error Rate Simulation

### 2.1. SEU Cross Section Angular Averaging

The key concept in single event effects is the sensitive volume [4] which, due to the non-locality of the impact of individual particles, is a single thin layer with an area of IC' memory and an effective thickness of $t_{eff}$ [5]. This effective thickness turns out to be quite small (of order tens nanometers) for ICs with low critical charge $Q_C$. This critical charge is the most important circuit parameter characterizing the noise immunity of digital elements, including to individual ionizing particles.

Based on statistical consideration we found that the SEU cross section (probability) per bit can be explicitly estimated by the value of the collected charge $\Delta Q$ or the energy $\Delta E \cong 22.5 (\Delta Q / \text{f C})$ keV deposited in sensitive volume during the passage of single ionizing particle [6]

$$\sigma(\Delta E) = \frac{a_C}{\exp\left(\dfrac{\varepsilon_C}{\Delta E}\right) - 1}, \qquad (1)$$

where $a_C$ is the area of the memory cell, $\varepsilon_C$ is the critical energy corresponding to circuit parameter $Q_C$. The energy deposition generally depends on the angle of incidence of the ion relative to the normal to the surface of the IC

$$\Delta E(\theta) = \Lambda \rho_{Si} t_{eff} / \cos\theta, \qquad (2)$$

where, $\Lambda$ is the ion's linear energy transfer (LET), $\rho_{Si} \cong 2.33$ g/cm³ is the silicon mass density, $t_{eff}$ is the effective thickness of the thin sensitive volume. Then, the angular dependent SEU cross section can be rewritten as follows

$$\sigma(\Lambda,\theta) = \frac{a_C}{\exp\left(\frac{2\Lambda_C}{\Lambda}\cos\theta\right)-1} \cong \begin{cases} \frac{K_d}{\cos\theta}(\Lambda - \Lambda_C \cos\theta), & \Lambda > \Lambda_C, (a) \\ a_C \exp\left(-\frac{2\Lambda_C}{\Lambda}\cos\theta\right), & \Lambda < \Lambda_C, (b) \end{cases}, \qquad (3)$$

where the slope $K_d$ and threshold of the quasi-linear part of the curve are directly measured at normal ion incidence $\theta = 0°$ experimental parameters

$$K_d \cong \frac{a_C}{2\Lambda_C} \text{ (a)}, \qquad \Lambda_C = \frac{\varepsilon_C}{\rho_{Si} t_{eff}} \text{ (b)}. \qquad (4)$$

Note that Eq. 3 describes both the multiple cell (MCU, $\sigma \geq a_C$)[26] at high LETs and the single bit (SBU, $\sigma < a_C$)[27] upset modes at low LETs. It is also important to note that the term "critical LET" here refers to the result of unambiguous interpolation of the linear portion of the cross-section dependence on LET, obtained at normal ion incidence (as in Eq. 3a).

In space we have an approximately isotropic particle flow, which implies averaging of the SEU cross-section of failures over the full solid angle

$$\langle\sigma(\Lambda)\rangle = \int \sigma(\Lambda,\theta)|\cos\theta|d\cos\theta. \qquad (5)$$

Using (5) and (3a) one can get

$$\langle\sigma(\Lambda)\rangle \cong 2K_d\left(\Lambda - \frac{\Lambda_C}{2}\right), \qquad \Lambda > \Lambda_C. \qquad (6)$$

This corresponds to an average chord length equal to twice the thickness of the sensitive area $\langle t_{eff}(\theta)\rangle = 2t_{eff}$ (see App. 1b). Due to the influence of grazing incidence angles, the effective value of $K_d$ doubles and the effective value of $\Lambda_C$ is halved when averaging over all angles compared to the value measured at $\theta = 0°$ at $\Lambda > \Lambda_C$.

Next, averaging Eq.(3b), we obtain

$$\langle\sigma(\Lambda)\rangle = a_C \frac{\Lambda}{2\Lambda_C^2}\left(\Lambda - (\Lambda + 2\Lambda_C)e^{-2\frac{\Lambda_C}{\Lambda}}\right) \cong a_C \frac{\Lambda^2}{2\Lambda_C^2} = K_d \frac{\Lambda^2}{\Lambda_C}, \qquad \Lambda < \Lambda_C. \qquad (7)$$

So we have a good piecewise approximation, continuous at $\Lambda = \Lambda_C$ together with the first derivative (i.e. with no change in slope)

$$\langle\sigma(\Lambda)\rangle \cong \begin{cases} 2K_d(\Lambda - \Lambda_C/2), & \Lambda \geq \Lambda_C, \\ K_d \Lambda^2/\Lambda_C, & \Lambda \leq \Lambda_C, \end{cases} \qquad (8)$$

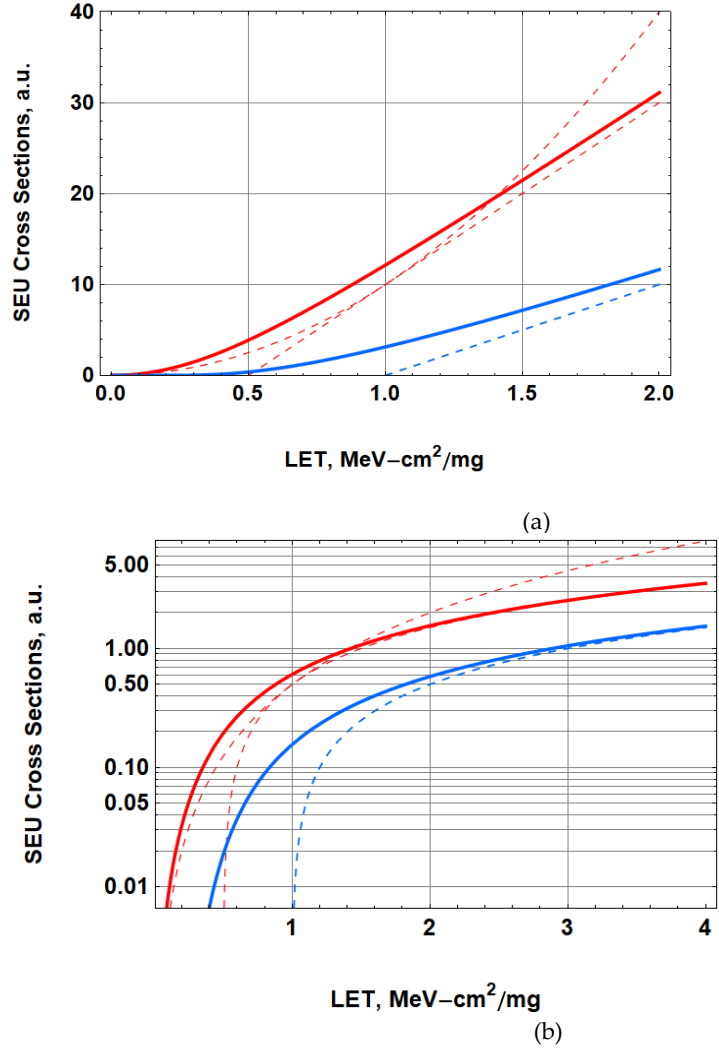

**Figure 1.** (a) The SEU cross section LET dependencies with before (blue) and after (red) angular averaging. Dashed lines show linear and square approximations at $\Lambda_c = 1$ MeV-cm²/mg (b) the same curves in logarithmic scale. Solid red lines show the results of exact analytical averaging (see Appendix A).

### 2.2. Soft error rate representations

For given orbital LET spectra, the soft error rate (SER) per bit can be expressed [8] in two equivalent forms

$$R = \int \langle \sigma(\Lambda) \rangle \Phi_d(\Lambda) d\Lambda = \int \Phi(>\Lambda) \frac{d\langle \sigma(\Lambda) \rangle}{d\Lambda} d\Lambda \qquad (9)$$

where $\Phi(>\Lambda)$ is the integral LET spectrum (i.e., the flux of particles with LET great than a given $\Lambda$), and $\Phi_d(\Lambda)$ is the differential LET spectrum, which are consistently defined as follows

$$\Phi(>\Lambda) = \int_\Lambda^\infty \Phi_d(\Lambda') d\Lambda', \qquad \Phi_d(\Lambda) = -d\Phi(>\Lambda)/d\Lambda \qquad (10)$$

### 2.4. LET specta parametrization

The galactic cosmic ray (GCR) flux (fluence per time) LET spectra can be parameterized by power law dependencies $\Phi_d(\Lambda) \propto \Lambda^{-n}$. In particular, it is known that the differential LET spectrum in the range up to 30 MeV×cm²/mg is well described at n=3 [9].

$$\Phi_d(\Lambda) \cong b\Lambda^{-3}, \quad \Phi(>\Lambda) \cong b/2\Lambda^2, \qquad (11)$$

The empirical constant $b$ characterizes the LET spectrum in a given orbit and depends on orbit's type, inclination and altitude, shielding and space weather.

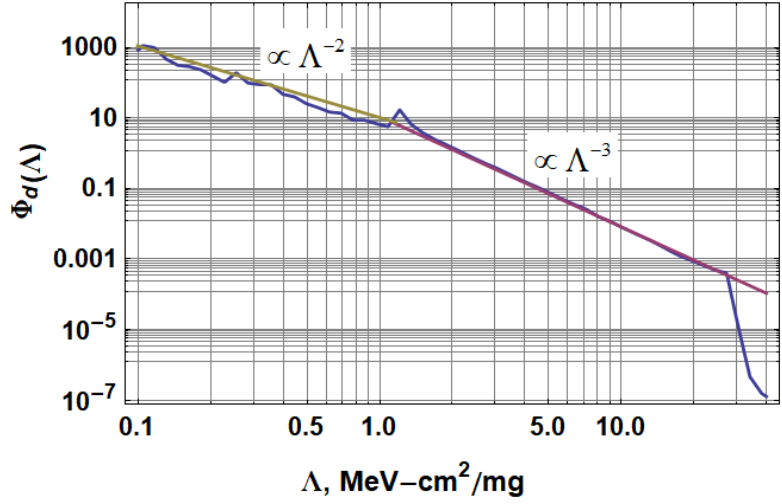

**Figure 2.** Approximation of the real LET spectrum (in $(cm^2 \times hour \times MeV\text{-}cm^2/mg)^{-1}$) on geostationary orbit using a power function.

Besides, the calculated LET spectra usually show a break at some value $\Lambda_r$ so that at $\Lambda < \Lambda_r$ the inverse cubic dependence is replaced by approximately inverse square dependence (see Fig.1). The, the parameter $b$ depends on the choice of the LET reference point $b = 2\Lambda_r^2 \Phi(>\Lambda_r)$ such that the differential and integral LET spectra can be conveniently parameterized at $\Lambda \geq \Lambda_r$ as follows

$$\Phi_d(\Lambda) \cong \frac{2}{\Lambda_r}\left(\frac{\Lambda_r}{\Lambda}\right)^3 \Phi(>\Lambda_r), \qquad \Phi(>\Lambda) = \left(\frac{\Lambda_r}{\Lambda}\right)^2 \Phi(>\Lambda_r), \qquad \Lambda \geq \Lambda_r. \quad (12)$$

On the other hand, with $\Lambda < \Lambda_r$ we have

$$\Phi_d(\Lambda) = \frac{\Lambda_r}{\Lambda^2}\Phi(>\Lambda_r), \qquad \Phi(>\Lambda) = \frac{\Lambda_r}{\Lambda}\Phi(>\Lambda_r), \qquad \Lambda < \Lambda_r, \quad (13)$$

where $\Phi(>\Lambda_r)$ is not a function but simply a number characterizing the integral fluence of particle, say, per hour in a given orbit and for given conditions.

The integral spectra (12) and (13) are continuous at $\Lambda = \Lambda_r$, and the differential spectrum experiences a jump in slopes at this point, as can be seen in Fig. 2. In addition, the spectra at $\Lambda \leq \Lambda_r$ are very irregular in nature, which is explained by both physical and computational uncertainties and problems. For this reason, the smoothed approximation is a necessary step to obtain analytical assessments.

### 3. SER analytical formulas
#### 3.1. Piecewise representation of SER

Depending on the values of the parameters $\Lambda_C$ and $\Lambda_r$, we have some combinatorics of the contributions of different integrals to the total value of SER. Provided $\Lambda_C \geq \Lambda_r$ we have 3 independent integration intervals: $0 < \Lambda < \Lambda_r$, $\Lambda_r < \Lambda < \Lambda_C$, and $\Lambda_C < \Lambda$. Using Egs. 8 and 13-14 we calculate the contribution of each LET interval

$$R(>\Lambda_C) = \int_{\Lambda_C}^{\infty} 2K_d \left(\Lambda - \frac{\Lambda_C}{2}\right) \frac{b}{\Lambda^3} d\Lambda = b \frac{3K_d}{2\Lambda_C} =$$
$$= 3K_d \Lambda_C \left(\frac{\Lambda_r}{\Lambda_C}\right)^2 \Phi(>\Lambda_r);$$
(14)

$$R(\Lambda_r < \Lambda < \Lambda_C) = \frac{K_d}{\Lambda_C} \int_{\Lambda_r}^{\Lambda_C} \Lambda^2 \frac{b}{\Lambda^3} d\Lambda = \frac{K_d}{\Lambda_C} b \ln\left(\frac{\Lambda_C}{\Lambda_r}\right) =$$
$$= 2K_d \Lambda_C \left(\frac{\Lambda_r}{\Lambda_C}\right)^2 \ln\left(\frac{\Lambda_C}{\Lambda_r}\right) \Phi(>\Lambda_r);$$
(15)

$$R(\Lambda < \Lambda_r) \cong \frac{K_d}{\Lambda_C} \int_0^{\Lambda_r} \Lambda^2 \frac{\Lambda_r}{\Lambda^2} \Phi(>\Lambda_r) d\Lambda = K_d \Lambda_C \left(\frac{\Lambda_r}{\Lambda_C}\right)^2 \Phi(>\Lambda_r).$$
(16)

Then the total SER at $\Lambda_C \geq \Lambda_r$ is represented by the sum of Eqs. 14-16

$$R_{tot} = K_d \Lambda_C \left(\frac{\Lambda_r}{\Lambda_C}\right)^2 \Phi(>\Lambda_r) \left(4 + 2\ln\left(\frac{\Lambda_C}{\Lambda_r}\right)\right), \quad \Lambda_C \geq \Lambda_r.$$
(17)

On another condition $\Lambda_C \leq \Lambda_r$ we also have 3 independent integration intervals: $0 < \Lambda < \Lambda_C$, $\Lambda_C < \Lambda < \Lambda_r$, $\Lambda_r < \Lambda$, with three contribution integrals:

$$R(0 < \Lambda < \Lambda_C) = \frac{K_d}{\Lambda_C} \int_0^{\Lambda_C} \Lambda^2 \frac{\Lambda_r}{\Lambda^2} \Phi(>\Lambda_r) d\Lambda = K_d \Lambda_r \Phi(>\Lambda_r) =$$
$$= K_d \Lambda_C \frac{\Lambda_r}{\Lambda_C} \Phi(>\Lambda_r)$$
(18)

$$R(\Lambda_C < \Lambda < \Lambda_r) = \int_{\Lambda_C}^{\Lambda_r} 2K_d \left(\Lambda - \frac{\Lambda_C}{2}\right) \frac{\Lambda_r}{\Lambda^2} \Phi(>\Lambda_r) d\Lambda =$$
$$= K_d \left(\Lambda_r - \Lambda_C + 2\Lambda_r \ln\left(\frac{\Lambda_r}{\Lambda_C}\right)\right) \Phi(>\Lambda_r)$$
(19)

$$R(>\Lambda_r) \cong \int_{\Lambda_r}^{\infty} 2K_d \left(\Lambda - \frac{\Lambda_C}{2}\right) \frac{b}{\Lambda^3} d\Lambda = K_d (4\Lambda_r - \Lambda_C) \Phi(>\Lambda_r)$$
(20)

Then, the total SER is the sum of the contributions of different regions of LET spectra

$$R_{tot} = K_d \Lambda_C \Phi(>\Lambda_r) \begin{cases} \left(\frac{\Lambda_r}{\Lambda_C}\right)^2 (4 + 2\ln(\Lambda_C/\Lambda_r)), & \Lambda_C \geq \Lambda_r, \\ 6\frac{\Lambda_r}{\Lambda_C} + 2\frac{\Lambda_r}{\Lambda_C} \ln\left(\frac{\Lambda_r}{\Lambda_C}\right) - 2, & \Lambda_C \leq \Lambda_r. \end{cases}$$
(22)

Fig. 3 shows SER as a function of critical LET calculated with Eq. 22 compared to the well-known Petersen Figure-of-Merit (FOM) model, in which SER $\propto \sigma_{SAT}/\Lambda_{th}^2$, where $\Lambda_{th}$ ("threshold" LET) is a parameter similar to the $\Lambda_C$, but is extracted from Weibull interpolation.

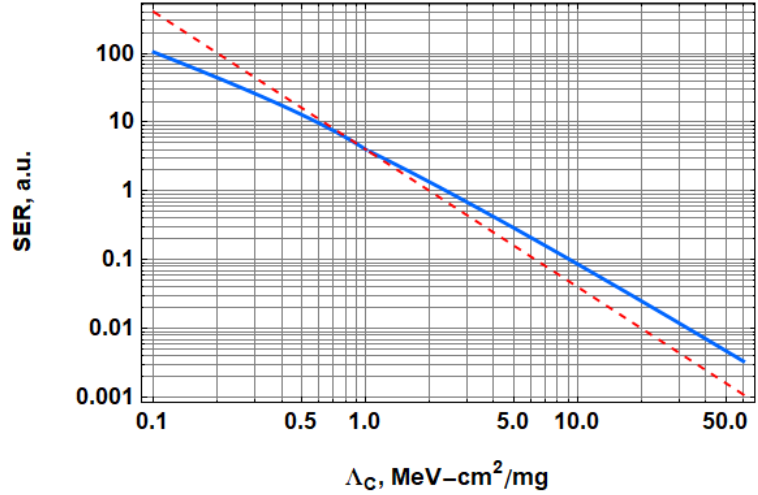

**Figure 3.** (a) The SER vs critical LET at $\Lambda_r$ = 1 MeV-cm²/mg. Dashed line shows the Petersen FOM dependence $\propto 1/\Lambda_C^2$. For ease of comparison, the absolute values of the calculations for the two approaches are equated at 1 MeV-cm²/mg.

Let us recall that the key point of the Petersen's approach was that the cross section vs LET dependence is approximated through the step function $\theta(x)$ as follows

$$\sigma(\Lambda) = \sigma_{SAT} \theta(\Lambda - \Lambda_{th}) \qquad (23)$$

where $\sigma_{SAT}$ ("saturation" cross section) and $\Lambda_{th}$ ("threshold" LET) are parameters of Weibull approximation. Then the integral (10) directly gives an expression like

$$R_{Petersen} = \frac{b\,\sigma_{SAT}}{2\Lambda_{th}^2} = \sigma_{SAT} \Phi(>\Lambda_{th}) \qquad (24)$$

The main problem with this approach is that the values of $\sigma_{SAT}$ and $\Lambda_{th}$ are poorly defined experimentally. In particular, the experimental SEU cross section does not typically saturate with increasing LET and $\sigma_{SAT}$ depends on the maximum LET value used in the test. In turn, $\Lambda_{th}$ depends on the minimum LET value to be used.

An aggravating problem is that $\sigma_{SAT}$ and $\Lambda_{th}$ are extracted from the experimental data as parameters of the Weibull function. It is shown [5] that this is fundamentally impossible to do unambiguously because of the mathematical structure of the distribution and due the presence of two additional physically meaningless fitting parameters in Weibull function. As can be seen from Fig. 3, the Petersen's simplified approach overestimates SER at low $\Lambda_C$ and underestimates it at high $\Lambda_C$.

### 3.1. Effective cross section

It is useful to define some effective value of the cross section independent of the orbital spectrum parameters. It seems reasonable to normalize the orbit depndent SER to the total fluence of particles with LET greater than some reference value $\Lambda > \Lambda_r$. Then the effective cross section can be defined as follows

$$\sigma_{eff} = \frac{\int \langle \sigma(\Lambda) \rangle \Phi_d(\Lambda) d\Lambda}{\Phi(\Lambda > \Lambda_r)} \qquad (25)$$

Since the LET spectra for different orbits differ more by the scaling factor $b$ rather than the shape of the spectrum $\Phi_d(\Lambda) \cong b\,f(\Lambda)$, the effective cross section is almost orbit-independent and is a characteristic of the integrated circuit rather than the space environment. Thus, taking into account (22) and (25) the orbit independent effective cross section can be written as follows

$$\sigma_{eff} = a_C \begin{cases} \left(\dfrac{\Lambda_r}{\Lambda_C}\right)^2 \left(2 + \ln(\Lambda_C/\Lambda_r)\right), & \Lambda_C \geq \Lambda_r, \\ 3\dfrac{\Lambda_r}{\Lambda_C} + \dfrac{\Lambda_r}{\Lambda_C}\ln\left(\dfrac{\Lambda_r}{\Lambda_C}\right) - 1, & \Lambda_C \leq \Lambda_r. \end{cases} \quad (26)$$

Fig. 4 shows a simulated with Eq. 26 contour plot of SER (effective SEU cross section) dependence both on the critical LET $\Lambda_C$ and the memory cell area $a_C$.

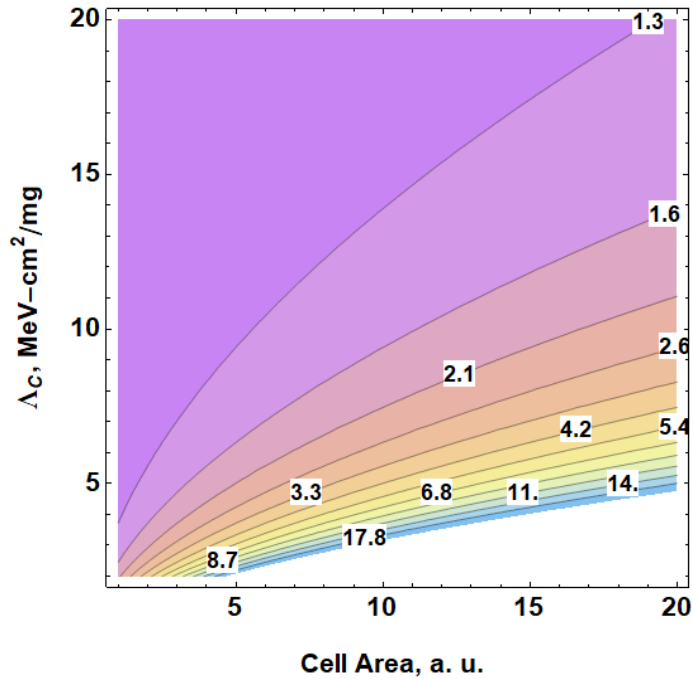

**Figure 4.** Parametric contour plot of the SER in arbitrary units as function of the critical LET and the memory cell area.

The lines in Fig. 4 correspond to equal values of SER (or, effective SEU cross section) in arbitrary units simulated with Eq. 26. Miniaturization of components within the framework of geometric scaling generally leads to a decrease in both the area $a_C$ and the critical LET (energy) $\Lambda_C$. However, decreasing $\Lambda_C$ results in an increase in SER, while decreasing $a_C$ suppresses SER due to a decrease in the probability of ion strike. This circumstance allows for non-monotonic behavior of SER under aggressive scaling [10].

## Conclusion

Based on the analytical averaging of the SEU angular characteristics, an explicit expression for the SER of digital elements in space is obtained, which has the form of the product of the effective cross section and the integral particle flux in a given orbit. The effective cross section per bit does not depend on the orbit parameters and is determined only by the experimentally determined value of the critical LET measured at normal incidence (closely related to the critical charge) and by the area of the memory cell.

## Appendix A

The exact value of the integral in Eq. 5 can be represented by the general formula

$$\dfrac{\langle \sigma(\Lambda) \rangle}{a_C} = \dfrac{\Lambda^2}{2\Lambda_C^2}\left(Li_2\left[e^{2\frac{\Lambda_C}{\Lambda}}\right] - \dfrac{\pi^2}{6}\right) + \dfrac{\Lambda}{\Lambda_C}\ln\left(1 - e^{2\frac{\Lambda_C}{\Lambda}}\right) - 1 \quad (A1)$$

where $Li_2[z]$ is the dilogarithm function [11]. Equation (A1) is a real-valued function for any LET and can be computed explicitly using standard functions in Python or Mathematica, for example. The results of calculations using this formula are presented in Figs. 1 as solid red lines. However, the exact formula has a cumbersome and not entirely transparent appearance, while differing little from a simple two-piece approximation.

**Appendix B**

The track length in a layer of thickness $t_{eff}$, much smaller than the seizes of the sensitive area, is described as

$$s = \frac{t_{eff}}{\cos\theta}. \tag{B1}$$

Then, the averaging over the angular distribution for an isotropic flow yields

$$\langle s \rangle = \int \frac{t_{eff}}{\cos\theta}|\cos\theta|d\cos\theta = 2\int_0^{\pi/2} \frac{t_{eff}}{\cos\theta}\cos\theta\sin\theta\,d\theta = 2t_{eff}. \tag{B2}$$

This is equivalent to averaging over the angular distribution for an isotropic flow

$$f(\theta) = 2\cos\theta\sin\theta d\theta. \tag{B2}$$

Taking into account $ds/d\theta = s\tan\theta$ the angular distribution can written as a track length distribution $f(\theta)d\theta = f(s(\theta))ds$, where the latter can be represented as follows

$$f(s) = \frac{2t_{eff}^2}{s^3}, \qquad F(s) = \frac{t_{eff}^2}{s^2}, \tag{B3}$$

where $F(s)$ is the integral chord length distribution. This gives the same averaging result

$$\int_{t_{eff}}^{\infty} f(s)s\,ds = 2t_{eff}. \tag{B4}$$

According to, for example, the SEU cross section per device, averaged over chord lengths, can be written as follows

$$\langle \sigma(\Lambda) \rangle = \frac{A_{tot}}{4} F\left(\frac{\varepsilon_C}{\rho_{Si}\Lambda}\right), \qquad \frac{\varepsilon_C}{\rho_{Si}\Lambda} < t_{eff}, \text{ or, } \Lambda > \Lambda_C, \tag{B5}$$

that is equivalent to the Bradford approach [11]. Here, $A_{tot}$ is the total surface area of the sensitive volume which for thin lamina is the twice memory area value $A_{tot} = 2A_M$. Then, taking into account Eqs. B3 we get the relation

$$\langle \sigma(\Lambda) \rangle = \frac{A_M}{2}\frac{\Lambda^2}{\Lambda_C^2} = N_{bit}K_d\frac{\Lambda^2}{\Lambda_C}, \qquad \Lambda > \Lambda_C \tag{B6}$$

where $A_M = N_{bit}a_C$ which exactly corresponds to the result of angular averaging in Eq. 7. Thus, we have verified that averaging over angles and over chord lengths gives equivalent results, and Eq. (3) can be considered as the result of microdosimetric averaging over the distribution of track lengths of a single sensitive region of thickness $t_{eff}$ and an area of the order of the IC sizes. It is fundamentally important that the equivalence between averaging over angles and track lengths holds only for a single sensitive volume unlike the so-called RPP method [12-14], in which averaging over chord lengths is performed for a separate memory cell. The idea that such averaging can be carried out independently is mathematically inconsistent and erroneous due to the fundamental non-locality of the effect of an individual ion [5]. Such non-locality holds even in large-sized cells of old technologies at grazing angles of incidence. One of the most important manifestations of such non-locality in modern ICs is the multiple cell upset effect. Ideologically, our method is closest to the so-called Effective Flux Model [15], which is also in fact based on averaging over angles. Nevertheless, unlike our approach, the RPP and Effective Flux models do not in principle describe multiple cell upsets.